# Near–field radiative heat transfer between a sphere and a substrate

Arvind Narayanaswamy

Department of Mechanical Engineering, Columbia University, New York, NY 10027.

Sheng Shen and Gang Chen

Department of Mechanical Engineering, Massachusetts Institute of Technology,

Cambridge, MA 02139.

**Near–field force and energy exchange between two objects due to quantum electrodynamic fluctuations give rise to interesting phenomena such as Casimir and van der Waals forces, and thermal radiative transfer exceeding Planck's theory of blackbody radiation. Although significant progress has been made in the past on the precise measurement of Casimir force related to zero-point energy, experimental demonstration of near-field enhancement of radiative heat transfer is difficult. In this work, we present a sensitive technique of measuring near–field radiative transfer between a microsphere and a substrate using a bi–material atomic force microscope (AFM) cantilever, resulting in "heat transfer-distance" curves. Measurements of radiative transfer between a sphere and a flat substrate show the presence of strong near–field effects resulting in enhancement of heat transfer over the predictions of the Planck blackbody radiation theory.**

Though Planck realized that the theory of blackbody radiation is applicable only to objects with characteristic dimensions larger than the wavelength of thermal radiation [1], a rigorous theory of near–field radiative transfer was established only later by Polder and van Hove [2] and others [3-8] following the fluctuational electrodynamics formalism established by Rytov [9]. The modification of thermal fluctuations of the electromagnetic field due proximity between two objects is the source of near-field thermal radiative transfer as well as thermal contributions to dispersion forces such as Casimir and van der Waals forces. Experimentally, there are a few reports [10-13] on near-field heat transfer experiments, but none has demonstrated exceeding Planck's blackbody radiation law [14]. Tien and co-workers reported increased near–field radiative heat transfer at cryogenic temperatures between two parallel metallic plates at gaps of 50 μm to 1 mm [10], but the total heat transfer is ≈ 1/29 of the Planck theory prediction. Hargreaves [11] extended the measurements, between two chromium surfaces, to gaps up to 1 μm at room temperature, demonstrating an enhancement from ≈ 1.75 Wm$^{-2}$K$^{-1}$ in the far–field to ≈ 2.95 Wm$^{-2}$K$^{-1}$ at 1 μm. However, the near-field radiation heat transfer between two chromium surfaces is still less than 50 % of blackbody radiation. Xu et al. [12] could not measure any signature of near-field enhancement because of the low sensitivity of their experimental technique. Recently, Kittel et al. [13] investigated the near–field radiative transfer between a sharpened scanning tunneling microscope tip and a flat substrate and showed a saturation of heat transfer at gaps of 10 nm or less and a decrease at larger gaps. The saturation effects and enhancement of heat transfer have been attributed to spatial dispersion effects and the contribution of the infrared magnetic dipole component [15, 16]. However, the complicated geometry of the tip makes it difficult to interpret the experimental data. More precise measurements are needed to confirm past extensive theoretical studies and predictions of near-field enhancement, especially that of exceeding Planck's theory of blackbody radiation. Such near-field experiment also provides insight into thermal contributions to the Casmir force, which has been elusive to experimental detection because of its much smaller magnitude compared to the contribution of the zero-point energy [17, 18].

We decided to measure radiative transfer between a microsphere and a flat substrate in order to overcome the difficulties encountered in the past experiments and use bi-material AFM cantilevers as thermal sensors. Bi-material cantilevers bend in response to changes in temperature distribution in the cantilever due to the difference in coefficient of thermal expansion of the two materials comprising the cantilever. They have been used as sensitive calorimeters [19, 20], and IR detectors [21, 22]. Such cantilevers are reported to have a minimum measurable temperature of $10^{-4}$ K to $10^{-5}$ K and minimum detectable power of $5 \times 10^{-10}$ W (when optimized, the minimum detectable levels are even lower) [20]. With a temperature difference of 50 K between the sphere and the substrate, the minimum detectable conductance is $\lfloor 10^{-2}$ W K$^{-1}$. A similar configuration has also been used for high precision measurements of Casimir force [17, 23]. Using the rigorous theory for near-field radiative transfer between two spheres that we have developed elsewhere [8], it is possible to estimate the values of thermal conductance between a sphere and a substrate [24]. This theory predicts that the conductance between two silica spheres of diameters 50 μm ranges from $\lfloor 10^{-2}$ W K$^{-1}$ to $\lfloor 10^{-2}$ W K$^{-1}$ for gaps between 100 nm and 10 μm, well within the measurement limits of the bi-material cantilever. The near-field enhancement is due to the tunneling of surface phonon polaritons present at interfaces between silica and vacuum due to thermal fluctuations of the electromagnetic field. Hence we have chosen silica spheres of diameter 50 μm for our experiment.

A schematic of the experimental apparatus is shown in Fig. 1(a) and a silica sphere is attached to the tip of a triangular SiN/Au cantilever (from BudgetSensors) as shown in Fig. 1(b). To decrease the influence of dispersion and electrostatic forces, the cantilever is oriented perpendicular to the substrate (90° ± 2°). The substrate, which is rigidly attached to the motion control stage, is a glass microscope slide. The apparatus is placed inside a vacuum chamber pumped down to ≈ 6.7 × 10$^{-3}$ Pa during the experiment. The laser beam is focussed at the tip of the cantilever and the reflected beam forms a spot on a position sensitive detector (PSD). The position of the spot and the laser power incident on the PSD are obtained from the PSD difference and sum signals respectively. The change in position of this spot is a measure of the deflection of the cantilever. A portion of the laser beam is absorbed by the cantilever and results in a temperature rise of the cantilever tip and sphere. The base of the cantilever, the substrate, and the rest of the apparatus are approximately at the same temperature. As the gap between the sphere and the substrate decreases, increased heat transfer between the sphere and the substrate results in a cooling of the cantilever. The resultant deflection is measured as a change in the PSD difference signal, resulting in a "heat transfer–distance" curve. When the pressure inside the vacuum chamber is less than 0.1 Pa, the dominant form of heat-transfer mechanism is radiative transfer. As the radiation view factor from the sphere to the surrounding is | 1, and the substrate is at the same temperature as the environment, there is no change in the far-field radiative heat transfer even when the cantilever approaches the substrate. Hence, what is measured in the heat transfer-distance curve is exclusively the near-field enhancement above the far-field radiative transfer between two objects predicted by Planck's theory of radiative transfer.

To determine the conductance of radiative transfer between the sphere and the flat substrate, three quantities are necessary: (1) the heat absorbed by the cantilever (from the laser), (2) the heat transfer between the sphere and substrate, and (3) the temperature of the sphere. The absorptivity of the cantilever, determined by measuring the radiant power in the incident, transmitted, and reflected laser beams using the sum signal of the PSD, is approximately 0.13. The deflection of the cantilever is converted to heat transfer between the sphere and the substrate by determining the sensitivity of the cantilever to power absorbed at the tip. This sensitivity is determined by varying the radiant power of the incident laser beam. Knowing the amount of absorbed power at the tip, the temperature of the sphere is determined if the conductance of the cantilever, G, is known. This conductance is determined from the ratio of the sensitivity of the cantilever to uniform temperature rise of the cantilever to the sensitivity of the cantilever to power absorbed at the tip, as this ratio equals 2G [20, 25]. For the cantilever used in this work, the sensitivity to heat transfer from the tip is measured to be $9.28 \times 10^4$ VW$^{-1}$ and the sensitivity to uniform temperature change is 0.839 VK$^{-1}$, resulting in a conductance of $4.52 \times 10^{-6}$ WK$^{-1}$. With an incident power of 1.66 mW (as measured by a power meter), the absorbed power is $\approx 0.21$ mW and the temperature rise of the tip, which is also the temperature rise of the sphere, is 46.5 K.

Near–field effects become noticeable when the gap is approximately 10 μm or less [8]. The raw data from one of the experiments is shown in Fig. 2a. Shown in the figure are two curves corresponding to the deflection signal (y axis to the left) and the sum signal (y axis to the right). As time increases along the x-axis, the gap between the sphere and the substrate decreases. Contact between the sphere and the substrate manifests itself as a large change in the PSD signals. Once the contact point is known, the position of the substrate can be converted to an equivalent gap. The sum signal curve is flat even as the gap decreases. This ensures that the signal that is measured is predominantly due to the near–field effect and not due to any spurious effects [26]. The data from 13 heat-transfer distance curves are shown in Fig. 2b. Each red diamond marker in Fig. 2b corresponds to a data point in a heat transfer-distance curve. The scatter in the experimental data is 0.44 nW/K and is primarily because of the vibrations induced by the turbomolecular pump. In addition, the error bar in the x-axis due to the positional accuracy of the translation stage should be 100 nm. To understand the experimental data, we point out that the heat transfer-distance curve measures only the near-field effect. Hence, the data in Fig. 2b corresponds to the increase in near–field radiative transfer from the value at » 9 μm.

We have used Mie theory [27] to calculate the emissivity of a silica sphere of radius 25 μm to be 0.97. Using this value of emissivity, the far-field conductance between the sphere and the substrate is approximately equal to the blackbody conductance between the sphere and the substrate. Hence the maximum measured conductance due to near-field enhancement, as seen in Fig. 2b, is 6 nWK$^{-1}$ above the prediction of Planck's theory of blackbody radiation value of 29 nWK$^{-1}$ [28]. From the analysis of near–field radiative transfer between two spheres, we see that the near-field conductance between the two spheres varies as $Ax^{-n}$, where $n$ is an exponent less than 1. If $n = 1$, then the proximity approximation, which is widely used in determining forces between smoothly curved surfaces based on the results of forces between parallel surfaces [29, 30], would be valid. Since the experimental measurements are relative to the gap from which measurements are started, the experimental data should be of the form $Ax^{-n} + B$, where $B < 0$. For the data shown in Fig. 2b, the values of $n$, $A$, and $B$ are 0.55, 2.061, and -0.7, in reasonable agreement with the numerical solution of the two-sphere which yields values of 0.41, 2.41, and -0.978 for $n$, $A$, and $B$ respectively [26]. It is also clear from Fig. 2b that the experimental data is larger than the predictions of the proximity approximation, pointing to the lack of validity of the proximity approximation for near-field radiative heat transfer. We have shown that a proximity approximation type theory is valid for those spheres where near-field effects dominate radiative transfer [8]. This condition is satisfied only by silica spheres of diameter less than ≈ 2 μm and clearly not valid for microspheres of diameter ≈ 50 μm.

In summary, we have introduced a sensitive technique based on bi-material cantilevers to investigate near–field radiative transfer, and report experimental data on radiative heat transfer between a silica sphere and a silica substrate. Our experimental technique is sensitive to near-field radiation alone. Our experimental data shows the breakdown of the Planck blackbody radiation law in the near-field, and also shows that proximity approximation cannot be applied to near-field radiation in the range of gaps involved in the experiment. Improvements in the current experimental setup [26] can yield better measurements, enabling us to investigate near–field heat transfer between spheres of smaller diameter as well as between conducting spheres. The present experiment should also be of great interest for probing the temperature dependent behaviour of the Casimir force.

This work is supported by DOE (DE-FG02-02ER45977) and ONR (N00014-03-1-0835). The authors would also like to thank Mr. Lu Hu for calculating the emissivity of silica spheres.

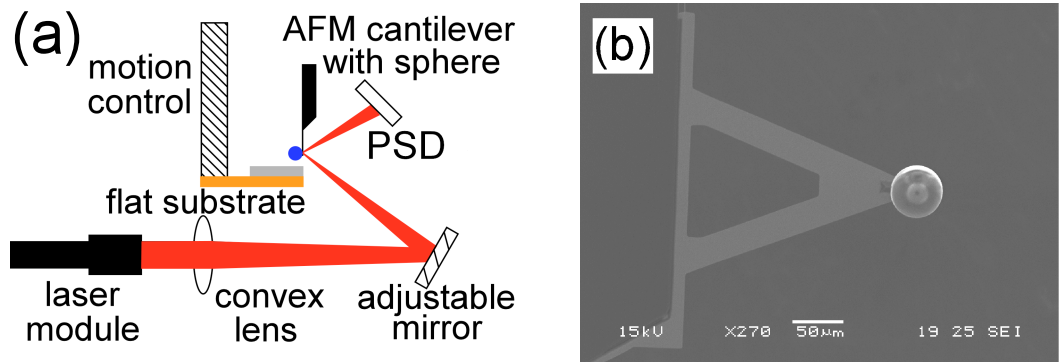

**FIGURE 1 (a) Schematic of experimental apparatus. The beam from a laser diode module is focussed at the tip of an AFM cantilever and the reflected portion is directed onto a PSD. The beam from the adjustable mirror onto the cantilever is the "incident beam", and the beam from the cantilever to the PSD is the "reflected beam". (b) SEM image of silica sphere attached to the tip of a triangular bi-material cantilever.**

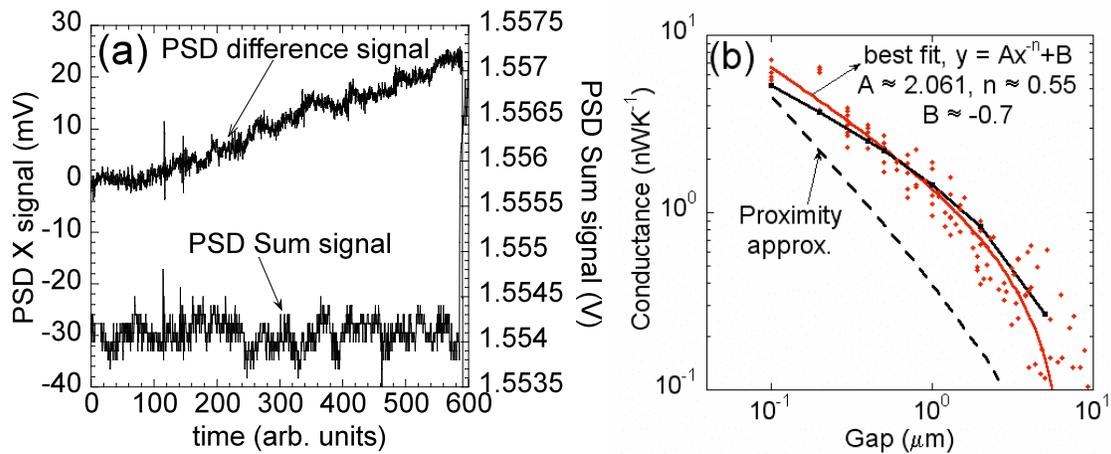

**FIGURE 2** (a) Raw experimental data from one of the experiments. The y–axis to the left corresponds to the PSD difference signal and that on the right corresponds to the PSD sum signal. The substrate is brought closer to the sphere (smallest step size is 100 nm) as the experiment proceeds. The contact is seen as a large change in the PSD signals. The sum signal is approximately a constant and this ensures that the deflection of the cantilever is not because of any spurious effect related to a change in the incident radiant power. (b) Experimental data (diamonds) from 13 heat transfer-distance measurements. The red line through the data is a best fit curve of the form $y = Ax^{-n} + B$. Also shown in the figure are the predictions of the proximity approximation (black dashed line) and the two-sphere problem [8, 26] predictions for the near-field transfer between a sphere and a flat plate (black line with black squares), obtained by multiplying the results of the two-sphere problem by 2 (The factor of 2 is chosen because the conductance between a sphere and a flat substrate at a given gap is twice the conductance between two spheres of the same radii and gap). The experimental configuration described in the text automatically measures the enhancement in radiative transfer due to near-field and diffraction effects, which are not included in Planck's theory of blackbody radiation, relative to the value at ≈ 9 μm.